# Probery: A Probability-based Incomplete Query Optimization for Big Data

Jie SONG, Yichuan ZHANG, Yubin BAO, Ge YU

**Abstract**— Nowadays, query optimization has been highly concerned in big data management, especially in NoSQL databases. Approximate queries boost query performance by loss of accuracy, for example, sampling approaches trade off query completeness for efficiency. Different from them, we propose an uncertainty of query completeness, called Probability of query Completeness (PC for short). PC refers to the possibility that query results contain all satisfied records. For example PC=0.95, it guarantees that there are no more than 5 incomplete queries among 100 ones, but not guarantees how incomplete they are. We trade off PC for query performance, and experiments show that a small loss of PC doubles query performance. The proposed Probery (PROBability-based data quERY) adopts the uncertainty of query completeness to accelerate OLTP queries. This paper illustrates the data and probability models, the probability based data placement and query processing, and the Apache Drill-based implementation of Probery. In experiments, we first prove that the percentage of complete queries is larger than the given PC confidence for various cases, namely that the PC guarantee is validate. Then Probery is compared with Drill, Impala and Hive in terms of query performance. The results indicate that Drill-based Probery performs as fast as Drill with complete query, while averagely 1.8x, 1.3x and 1.6x faster than Drill, Impala and Hive with possible complete query, respectively.

**Index Terms**— Big Data; NoSQL Database; Query Optimization; Probability; Query Completeness; Confidence

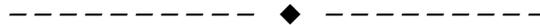

———————————— ◆ ————————————

## 1 INTRODUCTION

In the field of big data management, newly emerged NoSQL databases provide a mechanism for storage and management of data rather than the tabular relations used in relational databases [1]. The data structures used by NoSQL databases, such as key-value, wide column, graph, and document, are more flexible than relational database tables, making some operations faster [2]. Due to the huge volume of data, one of the main issues about query processing in NoSQL database is query performance, and both industry and academic concern the corresponding researches on its optimization. For big data queries, the performance is improved mainly by optimization approaches on tasks scheduling, query algorithm and data structure, or alternatively by approximate query processing. The former is mostly for accurate query processing, and the latter obtains approximate results with negligible deviation from the exact results, while it saves processing resources and dramatically improves the query performance.

Approximate query processing is suitable for big data. In many cases, it is impossible or too expensive for users to retrieve exact answers immediately, while an approximate query processing is an alternative way to answer users within the short response-time and without obvious errors. For example, in a Business intelligence (BI) applications, when we count the number of unique customer sessions on a website or establish the median house price within each zip code across a state, users accept the approximate results with quickly response rather than accurate results with huge delay.

An approximate query processing adopts precise aggregations on incomplete query results such as sampling approaches [3], or approximate algorithms such as approximate query conditions and approximate similarities [4]. Different from the existing studies of the approximate query processing, in this paper, we propose a probability-based query processing. It focuses on the uncertainty of query completeness, namely the possibility of query results containing all satisfied records, but not on the approximation of query results, namely the extent of query results closing to the true ones. We explain such uncertainty as "how confident the result set is complete" for a range query, and "how confident the result is aggregated from a complete result set" for an aggregated query. We trade off such uncertainty for query performance. Comparing with the existing studies, we believe the following reasons would support our idea:

(1) As we know, both partition and index techniques reduce scan scope and boost query efficiency well. These techniques base on a predefined data schema and a well-designed data structure, however, maintaining them is costly due to the "volume" and "variety" of big data.

(2) For an OLTP query, both performance and accuracy are more critical than completeness. Taking a range query as an example, if someone queries preferred restaurants nearby an airport, he may accept that there are possible missing restaurants that should match the query condition, or he even do not know there are the missing ones. Nevertheless, he suffers from the unmatched restaurants showing in the result set, or a huge response delay. Taking

———————————————

• *Jie Song, Ph.D., Associate Professor, Software College, Northeastern University. E-mail: songjie@mail.neu.edu.cn.*
• *Yichuan Zhang, Ph.D., Lecture, Software College, Northeastern University. E-mail: zhangyc@swc.neu.edu.cn.*
• *Yubin Bao, Ph.D. Professor, School of Computer Science and Engineering, Northeastern University. E-mail: baoyb@ mail.neu.edu.cn.*
• *Ge Yu, Ph.D., Professor, School of Computer Science and Engineering, Northeastern University. E-mail:yuge@mail.neu.edu.cn*

an aggregated query as an example, if someone queries the average cost of preferred restaurants nearby an airport, he may accept the fast response with the uncertainty caused by some matched restaurants not being counted in, but not the other ways around, such as unexpected restaurants being counted in, or response delay.

(3) In approximation query processing, the query accuracy is defined as the similarity between the query results and the true ones. However, on one hand, it is undetermined because the true results are unknown without re-querying; on the other hand, it is nonuniform because different queries have different presentations of errors. Treating the uncertainty of query completeness as the criterion of approximation, it is determined because it is pre-calculated by the probability model and is uniform because the probability is the proper presentation for any queries.

(4) In a sampling-based approximation query, the performance improvement is dominated by the sampling rate, but not the accuracy. The accuracy relates to many conditions such as the sampling randomness and the data distribution. However, using the uncertainty of query completeness as the criterion, we build a determinable relationship between performance and accuracy.

In this paper, we propose Probery (PROBability-based data quERY), a key-value oriented query optimization for big data. It adopts the uncertainty of query completeness to accelerate OLTP queries. Probery reduces the scan scope to improve the query performance, employs the concept of Probability of query Completeness (PC for short) to measure the uncertainty, and trades off PC for query performance. In Probery, PC is a probability that query results are complete or aggregated from a complete result set. There are four core concepts in Probery: cell, a logic data container; block, a physical data container; Probability of Existence (PE for short), the probability of a cell storing on a block; confidence, a PC guarantee. Given a query condition with a confidence, Probery knows the PEs of matched cells, thus it cuts down the queried blocks, improves query performance, and ensures the PC is larger than confidence. For example, if confidence =1, all blocks are queried, while if confidence=0.5, less blocks are queried. Notice that in such situation: confidence is not sampling rate, and NOT exact 50% blocks are queried. Probery does not guarantee how many blocks are skipped because the PE distribution is not uniform. In contrast, Probery guarantees that there are less than 50% incomplete results if the query is executed many times, but not guarantees how incomplete they are. To design Probery, the following issues are studied.

(1) Clarify the difference and relationships with "the query performance", "the extent of query completeness" and "the probability of query completeness (PC)".

(2) Define the existence probability model, in which the probability of a data block containing the data is called Probability of Existence (PE).

(3) For data with a same key, how to design PE distribution on data blocks, and how to implement PE using the data placement algorithm.

(4) For data with a same key, the data matched query conditions have difference PEs on different data blocks because of the non-uniform PE distribution. Thus, sizes of data blocks are unbalance. For all keys, how to implement the balance of data blocks using a predefined data arrangement.

(5) The query processing algorithm satisfy both PC requirement and the randomicity. The latter means that the algorithm should select blocks randomly if the query is executed many times with the same confidence. Without the randomicity, two unexpected situations are happened when confidence≠1, first, if a block with small PE really contains matched data, then it will always be skipped, and not query is complete, second, the incomplete result sets are always the same。

(6) The Drill based system architecture and implementation.

Probery is a novel query optimization approach for key-value data. Comparing with the other partitions or indexes based query optimization, Probery maintains partitions on the logic data model, and indexes the logic partitions to the physical ones with their PE. However, it is different with the traditional partition and index techniques for the following reasons:

(1) Probery reduces scan scopes by lossing PC, while partition and index techniques reduce scan scopes by refining them.

(2) Traditional partition techniques face the challenge of flexibility. However, Probery adopts a schmea-independed partition strategy, namely that the PE distribution satisfies a statical and immutable distribution. Our approach is more flexible.

(3) Traditional index techniques face the challenges of huge storage and searching cost in big data environments. However, Probery do not rely on any pre-computation and materialized techniques, thus the additional storage cost is zero.

Probery is an optimization approach but not a "SQL on Hadoop" system. Its implementation relies on a host system, such as Drill in this paper, however, it only updates the scan mechanism of the host system, and it inherits the performance features of the host system. In the paper, we design a series of experiments to validate proposed approaches and compare query performance of Probery with popular "SQL on Hadoop" systems, for example Drill [5] as the host, Impala [6] as the in-memory optimization, and Hive on Tez [7,8] as an index optimization. The results show that Probery meet the design purposes:

(1) The percentage of complete queries is larger than the given confidence for various cases.

(2) The additional cost of data loading is slight and acceptable.

(3) The small loss of PC doubles query performance, while the query performance remains unchanged if confidence is set to be 1, comparing with the host system.

(4) Optimization on scan operation is efficient and better than in-memory optimization, such as Impala, and in-line index optimization, such as Hive.

The rest of paper is organized as follows. Section 2 introduces the related works. Section 3 defines the logic and physical data models, and the probability model. Section 4 explains how the data is partitioned and placed to the blocks. Section 5 describes query processing approaches and error estimation. Section 6 introduces the system architecture of Probery and explains its components. Section 7 validates the data placement in loading, query completeness and query efficiency, and then compares query performance of Probery with Drill, Impala and Hive. Finally, conclusions and future works are summarized in section 8.

## 2 RELATED WORKS

The problem of effectively and efficiently querying data in a probabilistic way has recently gained a great deal of attention from the database research community, due to the two reasons: Frist, the wide spread of application scenarios where uncertain or incomplete data occur; Second, the imprecise or incomplete query results trading for performance optimization in big data. It would be clear enough the former issue plays a main role in this area but the latter one is also critical. To the best of our knowledge, Probery is the only solution for incomplete queries where the user could set to what extent are queries possible to be complete, gracefully trading PC for faster response time. Similarly, BlinkDB [9] trades off result accuracy for faster response time by running a (precise) query on an appropriate data sample. Merlin [10] trades off accuracy of estimated probabilities of approximate queries for faster response time. We category our work as "incomplete queries" against the queries over "incomplete data" [11], while BlinkDB [9] and Merlin [10] are researches on "approximate queries" against the query over "approximate data". So we categories the related works into four groups: query relaxation that is close to incomplete query, query over incomplete data, approximate query, and query over uncertain (approximate) data.

**(1) Query relaxation**

Interactive Data Exploration (IDE) is a key ingredient of a diverse set of discovery-oriented applications, including ones from scientific computing and evidence-based medicine [12]. In this application, data discovery is a highly ad hoc interactive process where users execute numerous exploration queries using varying predicates aiming to balance the trade-off between collecting all relevant information and reducing the size of returned data [13]. Query relaxation enables the application to execute the most restrictive version of a query first, progressively relaxing the query until the required number of hits are obtained [14]. It is a kind of way to adjust query completeness by flexible query conditions. However, the cost is performance. The known works are interactive query relaxation [15] and approximate query relaxation [16].

An alternative approach for improving the response time of exploratory queries is to present approximate results. Such as online processing techniques [17] and the related CONTROL project [18]. These techniques offer approximate answers, and their goal is to allow users to get a quick sense of whether query reveals anything interesting about the data. Probery could work in a way similar to IDE. For example, when users submit a query with a very small confidence and immediately get the results, they take a glance of results, and re-submit same query with a larger confidence if the previous one contains interesting but too few results. Meanwhile, even that Probery is for OLTP query, it supports OLAP queries as described in [19] by aggregated query. It also supports the capabilities of "discover data" and "analyze data" by further processing on possible complete result set.

Probery is contrary to query relaxation because query results with larger confidence are more comprehensive than results with smaller confidence. The crucial assumptions underlying query relaxation approaches are the users desire the complete queries and they can explicitly state whether the query is complete or not. In Probery, users cannot state the complete results and they desire guarantees about the completeness and latency before querying.

**(2) Query over incomplete data**

Incomplete databases, databases with missing data, are presented in many research domains. Research works concentrate on dealing with incomplete data with incomplete data models, incomplete data indexes, incomplete skyline and similarity search, etc. [20], while Probery concentrates on possible incomplete queries. There are three main handling methods for processing incomplete data, i.e., simple deletion method, missing value imputation approach, and special treatment method. Among them, the former two convert the incomplete data set to the complete one. In contrast, the latter tackles incomplete datasets directly, such as the new language ISQL [21], the new dominance relationship definition on incomplete data [22], the distance function for incomplete objects [23], and the probability estimation on incomplete data [24]. In [24], the probability model is for estimating the similarity between the query and the data objects with missing dimensions. Probery does the similar way. We adopt a probability model to estimate the existence of the queried data objects and storage locations, so that both [24] and Probery provide the guarantee of query accuracy.

**(3) Approximate query**

An approximate query is characterized by its core algorithm, its error model and accuracy guarantee, the amount of work saved at runtime, and the amount of additional resources it requires in pre-computation. In recent studies on big data queries, approximate query is used to provide an optimized idea for traditional query methods in various situations, such as A/B testing, hypothesis testing, exploratory analytics [25], feature selection [26] and big data visualization [27,28]. In BlinkDB [9], authors present an approximate query processing framework for running interactive queries on large volumes of data. Soon after the commercialization of BlinkDB by Databricks [29], other Big Data query engines also started to add support for approximation features in their query engines, such as SnappyData [30] and Facebook's Presto [21].

An approximate query result is as useful as its accuracy guarantees. Quantifying or estimating the extent of error in approximate results not only with variance or bias, but also with probability of existence. Bias and variance are used for assessing the approximation quality of a numerical estimate. However, in case of a query result is not a single number, but a relation, the probability of existence of an output tuple is the probability with which it would also appear in the output of the exact query. The probability of existence is well studied in the context of probabilistic databases [32] and can be easily computed using bootstrap [33]. In Probery, we also define the probability of existence (PE, see definition 6) to evaluate the probability whether a block contains the data, and to determine whether a block is included in the scan scope. PE in Probery only overlaps the name of error estimation in approximate query, but not the concept.

Sampling is a straightforward approach for approximate queries, by which query results may not accurate but still representative. It has a long history in databases for more than two decades [34,35]. Users can perform sampling in pre-computation time or query time, and over query results, base tables, or entire database [36]. Among sampling methods, a probability sampling is the method in which every unit in the population has a chance (greater than zero) of being selected in the sample, and this probability could be accurately determined. In Probery, we also adopt the similar techniques in the data placement, that is, we select the target block to place data according to a certain probability. However, sampling based approximation query are orthogonal to our problem. By sampling, the query results are incomplete, and the probability determines to what extent the incompleteness is. In Probery, the query results might incomplete, the probability measures the confidence of query completeness, that it, to what extent the query results being completeness.

**(4) Query over uncertain data**

Uncertain data is data that contains noise that makes it deviate from the original values, and a probabilistic database is an uncertain database in which the data have associated probabilities [37]. There are two kinds of uncertainty called "membership uncertainty" (tuple-level) [32,38] and "value uncertainty" (attribute level) [39,40]. The former treats tuples as uncertain events capturing the belief that they belong to the database. The latter usually represents attributes with a discrete or continuous probability distribution depending upon the data domain. We learn the concept of membership uncertainty and associated the "placement-ship uncertainty" with data placement, i.e. the data placing to the storage location is uncertain.

Query processing on uncertain data includes range queries, threshold queries, skyline queries, top-k queries, nearest-neighbor queries, aggregate queries, and join queries. Among which the former two inspire Probery. A probabilistic range query tries to retrieve the results with the form of (*o*, *p*), where *p* is the probability of object *o* satisfying the query conditions [41]. Threshold query retrieves the objects qualifying a set of predicates with certain threshold guarantees. A range query processing in Probery includes two steps. Frist, Probery finds the logic container of query target denoted as *cell*, and then calculates (*b*, *p*) where *b* is a storage location called block and *p* is the probability of the *cell* stored on the block. This step is analogy to the probabilistic range query. Second, a heuristic blocks-selection algorithm (see section 5) is performed on blocks with a given confidence. This step is analogy to the threshold probability query.

## 3 MODELS

In this section, the data model and probability model of Probery are described. They are for probability-based data placement and query, which are introduced in next two sections

### 3.1 Data Model

The data model of Probery includes the logic part and the physical part. Probery follows the well-known key-values data model in the most key-value stores. A table contains many records, and each record consists of many key-value pairs. Probery calls keys as attributes. In the following sections, we only take one table with many attributes (keys) into consideration. For multiple tables, Probery re-uses the solution of a single table. The logic data model, called table space, is a multi-dimensional one whose dimensions are frequently queried attributes in the table. It is partitioned by cells. The physical data model includes blocks and trunk files, which store the records.

**Definition 1 Query Attribute.** According to the prior knowledge, the attributes, which are frequently appeared in the query conditions, are query attributes**.**

**Definition 2 Segment.** An attribute is divided into equal-frequency intervals according to its value range. These intervals are segments, also *empty* is a default segments with contains empty value.

Sometime equal-frequency intervals are difficult to concluded, but approximate equal-frequency ones are easily drawn using the data distribution.

**Definition 3 Table Space and Cells.** A table space is a multi-dimensional data model. A query attribute is modeled as a dimension, and its segments are modeled as non-hierarchal dimensional values on the dimension. A table space is partitioned into many none-overlapping and equal-size cells by dimensional values of each dimension. The granularity of cells is determined by the number of segments on each dimension, while the latter is dominated by the data volume and the size of physical addresses.

**Definition 4 Trunks and Blocks:** Trunks, or trunks files, are fixed size files in a distributed file system. Blocks are containers of these trunks. They are implemented as directories of a distributed file system.

In conclusion, cells are the logic containers of records. Records are categorized into cells. Trunks, which are files maintained by host file system, are storage of records. Blocks, the physical locations, are container of trunks. The number of cells dominates the number of blocks, and the data volume dominates the number of trunks in a block, called blocks size. The blocks size is almost equal due to the data balancing (see next section).

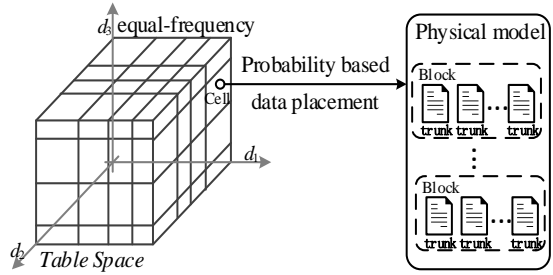

Fig. 1 the relationship among logical data model, block and physical data model

Fig. 1 shows an example of Probery data model. For the logic data model, the multi-dimensional table space has three dimensions $d_1$, $d_2$, $d_3$. Segments of these dimensions divide space into many cells. The number of records in each cell are almost same, and the range size of them vary because of uneven data distribution. For the physical data model, there are many blocks and trunks in each block. Probery maintains the mapping relationships among records, cells, blocks, trunks and nodes (machines). These mappings are list as the follows:

(1) Records to cells: many-to-one relationships are maintained using a table space.

(2) Records to blocks: one-to-one relationships are implemented by probability-based data placing. A record is possible to be placed to any blocks, however, it is finally placed in one block;

(3) Cells to blocks: many-to-many relationships are defined as an "existence", that is, a cell exists in a block if any record in the cell has been placed in the block.

(4) Trunks to blocks: many-to-one relationships are maintained by Probery, similar as files and folder. More trunks "maps" to a block if more records are placed in the block.

(5) Blocks to nodes: many to many relationships are implemented by distributed trunks of a block to all nodes uniformly.

(6) Trunks to nodes: many to one relationships are maintained by HDFS,

The probability model, which is adopted to build "mapping (2)" and calculated "mapping (3)", is described next.

### 3.2 Probability Model

Probability model consists the probabilities of query completeness, existence and placement. The first one defines the probability of query completeness. It is deduced by the second one, the probability of a cell exists in a block. The second one is accumulative probability of the third one, probability of a record be placed in a block.

**Definition 5 Probability of Query Completeness (PC):** PC is the probability that a query result is complete or being aggregated from a complete result set. In other word, it presents how confidence a query result set is complete.

PC is different with the Extent of query Completeness (EC for short). When a result set contains 90% of true results, EC is 0.9. However, if PC is 0.9, it means if a query be executed many times, 90% of cases whose ECs are 1, and 10% of cases whose ECs are not 1, but any values between 0 and 1.

**Definition 6 Probability of Placement (PP):** Given a record in a cell $c_i$ and a block $b_j$, the probability that the record is placed (stored) in $b_j$ is defined as the probability of placement. PP is associated with a cell and a block because all records in the cell have same PP to the block. Let PP of $c_i$ to $b_j$ be $p_{ij}$.

Let $m$ be the number of cells and $n$ be the number of blocks, PP is represented as an $m \times n$ matrix, where $p_{ij} = f_{DPA}(i,j)$. The function $f_{DPA}(i,j)$ is pre-defined, and the data placement is executed periodically according to the matrix. There will be explained in the next section.

**Definition 7 Probability of Existence (PE):** Given a cell $c_i$ of a table space and a block $b_j$, the probability that $b_j$ contains the records of $c_i$ is defined as the probability of existence. PE is associated with a cell and a block because the cell exists in the block if any record in the cell have been placed in the block. Let PE of $c_i$ to $b_j$ be $e_{ij}$. On the contrary, the Probability of Not Existence, denoted as PNE, is equals to 1-PE.

PE is represented as an $m \times n$ matrix where element $e_{ij}$ is the probability of $c_i$ existing in $b_j$. If $c_i$ has $\omega$ records, in other words, $c_i$ has been processed by data placement for $\omega$ times and one times for placing one record. Therefore, PNE of $c_i$ to $b_j$ is $(1-p_{ij})^\omega$, and $e_{ij} = 1-(1-p_{ij})^\omega$.

## 4 DATA PLACEMENT

Data placement is a method to place the records in cells into trunks in blocks. PC based querying relays on the data placement foundationally. In this section, data placement algorithm and PP distribution are discussed.

### 4.1 One cell

In this section, we explain how the records in one cell are placed according the PP distribution for the cell. The PP distribution is how PPs are distributed among all blocks, and denoted as $f(x)$.

**Function $h(x)$.** Let a continue function $h(x)$ satisfies:
(1) $h(x) \in (0,1)$;
(2) $\int_0^\lambda h(x)dx = 1-\varepsilon$, where $\lambda$ is a selected positive number, and $\varepsilon$ is a very small value.

Then, given a cell, $h(x)$ is modeled as probability density function of the cell. $H(x)$ is the primary function of $h(x)$.

**Function $f(x)$.** Given a cell, $f(x)$ is its PP to the block $x$. Let $n$ be the number of blocks, where $n > \lambda$ and $n\%\lambda = 0$, then let $\triangle = n/\lambda$. then numbering all blocks from 1 to $n$. the PP of block $a$ is $\int_{(a-1)\triangle}^{a\triangle} h(x)dx = H(\triangle a)-H(\triangle(a-1))$. Define $f(x) = H(\triangle x)-H(\triangle(x-1))$. When a record of the cell is loaded, Probery randomly generates a *seed* in the range of $f(x)$, then the record is placed in the block $\lfloor f^{-1}(seed) \rfloor$, or randomly select one of them if they are not unique.

**Function $g(x)$:** Given a cell, $g(x)$ is its PE to the block $x$. If there are $\omega$ records, after data placement, PE of the cell on the block $x$, denoted as $g(x)$, is $1-(1-f(x))^\omega$. We will explain the approach to avoiding extreme large $\omega$ value in section 4.3.

The definition of $g(x)$ is inspired by the Amdahl's Law that gives the theoretical speedup in latency of the execution of a task at fixed workload that is expected of a system whose resources are improved [42]. It predicts the theoretical speedup when using multiple processors. Comparing with a parallel-able task running in a standalone environment, when a few processors is introduced, the speedup is obvious; however, when the number of processors keeps on increasing, the speedup tends to be stable. It is explained that parallelism benefit the performance, however, it is impossible to get infinite parallelism by introducing more computers.

Probery is designed in the similar manger. The query performance is improved greatly if PC decreases a little, by many blocks with small PE being skipped in query; however, when PC decreases more greatly, the performance optimization effect is getting more stable, since few blocks with larger PE are skipped in the query. PE is the dominated attribute of probability-based query. PEs of blocks satisfy the rules as "less larger values" and "more small values". We select the normal distribution as $h(x)$ and deduce the functional image of $g(x)$, as shown in Fig 2.

(1) $h(x) = N(u, \sigma^2) = \dfrac{1}{\sqrt{2\pi\sigma^2}} e^{-\dfrac{(x-\mu)^2}{2\sigma^2}}$ where $\sigma$=0.3989.

(2) $H(x) = \dfrac{1}{2}\left[1 + erf\left(\dfrac{x-\mu}{\sigma\sqrt{2}}\right)\right]$

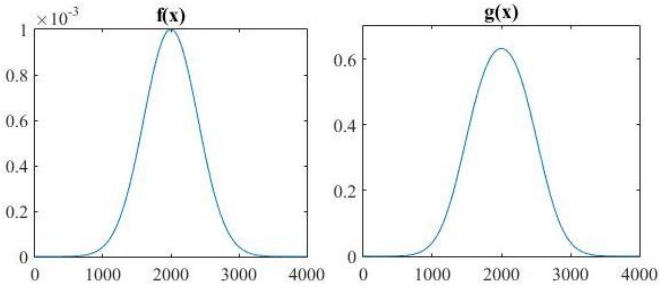

Fig 2 Image of $f(x)$ and $g(x) = 1-(1-f(x))^{\omega}$ when $\omega$ =1000, $\lambda$ =4, $n$=4000, $\sigma$=0.3989, $\mu$=0.5$\lambda$

## 4.2 Multiple cells

In this section, we extend the data placement of one cell to all cells. If $f(x)$ is applied to all cells directly, a block has the same PP for records of all cells, thus data distribution among blocks, which accords with $g(x)$, is unbalanced. Blocks are logic containers of trunks, and trunks of a block are uniformly stored among physical nodes. Therefore, we prefer a balance data distribution among blocks because it is benefit to parallelism.

**Definition 8. Data Placement Algorithm (DPA).** Data placement algorithm is a function $F_{DPA}: (i, j) \rightarrow p_{ij}$. Where $i \in [1,m]$ is the index of a cell in a table space, and $j \in [1,n]$ is the index of a block in the physical model, and $p_{ij}$ is probability of placement. The DPA satisfies the following three conditions.

(1) $\forall i \in [1,m]$, $F_{DPA}(i,1)+ F_{DPA}(i,2)+...+ F_{DPA}(i,n)=1$. That is, record has to be placed to a block.

(2) $\forall i \in [1,m]$, $j \in [1,n]$, $F_{DPA}(i,j) \neq 0$. That is, all blocks as the changes to be placed.

(3) $\forall a \in [1,m]$, $F_{DPA}(a,j)$ is continuous function with less large values and more small values. That is, data distribution of blocks is unbalanced for one cell.

(4) $\forall b_1 \in [1,n]$, $b_2 \in [1,n]$, $b_1 \neq b_2$ $\sum_{i=1}^{m} F_{DPA}(i,b_1) = \sum_{i=1}^{m} F_{DPA}(i,b_2)$, That is, data distribution of blocks is balanced for all cells.

Function $f(x)$ match the condition (1) to (3) for one cell, and an offset function is adopted for condition (4). Let offset be $\delta=n/m$, and

(1) $x=offset(i,j)= j+(i-1)\delta$ when $j+(i-1)\delta \leq n$ ;
(2) $x=offset(i,j)= j+(i-1)\delta-n$ when $j+(i-1)\delta > n$.

Then, $F_{DPA}(i, j)=f[offset(i,j)]$. For example, if $m$=3, $n$=6, $\delta$=2, then the $F_{DPA}$ and $f$ functions are mapped as Table 1. With the offset mechanism, the accumulative PP for all cells on any block is same, for example, the value is 0.25 in case of $m$=1000, $n$=4000, $\omega$=1000, $\lambda$ =4, $\sigma$=0.3989, $\mu$=0.5$\lambda$. Fig. 3 shows the PPs in gradient colors for all cells to all blocks in the case, and the symmetry of blocks is due to the offset. Fig. 3 prove both the condition (3) and (4) in Definition 8.

TABLE 1. EXAMPLE OF DATA PLACEMENT WHEN $M$=3 AND $N$=6

|  | Cell 1 | Cell 2 | Cell 3 |
|---|---|---|---|
| Block 1 | $F_{DPA}(1,1)=f(1)$ | $F_{DPA}(2,1)=f(3)$ | $F_{DPA}(3,1)=f(5)$ |
| Block 2 | $F_{DPA}(1,2)=f(2)$ | $F_{DPA}(2,2)=f(4)$ | $F_{DPA}(3,2)=f(6)$ |
| Block 3 | $F_{DPA}(1,3)=f(3)$ | $F_{DPA}(2,3)=f(5)$ | $F_{DPA}(3,3)=f(1)$ |
| Block 4 | $F_{DPA}(1,4)=f(4)$ | $F_{DPA}(2,4)=f(6)$ | $F_{DPA}(3,4)=f(2)$ |
| Block 5 | $F_{DPA}(1,5)=f(5)$ | $F_{DPA}(2,5)=f(1)$ | $F_{DPA}(3,5)=f(3)$ |
| Block 6 | $F_{DPA}(1,6)=f(6)$ | $F_{DPA}(2,6)=f(2)$ | $F_{DPA}(3,6)=f(4)$ |

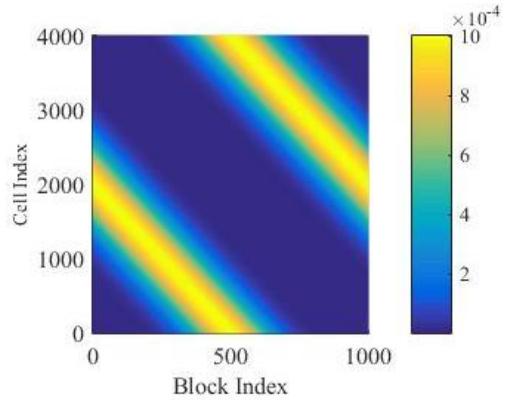

Fig 3 PPs for cells to blocks.

## 4.3 Practice experiences

In these section, we introduce some practice experiences for the implementation of Probery. These experiences do not change the algorithms and models of Probery, and they will improve the efficiency. We exclude these from the model and algorithm for keeping them simpler and more comprehensible.

In practice, to avoiding the situation that PE is either 0 or 1 for most blocks due to the extreme large $\omega$ value, we introduce the concept of slot. Slots are containers of blocks. If blocks are treated as folders, then slots are parent folders. Every slot contains $n$ block. The PP and PE distribution of blocks in each slot is independent and identical, that is, a block $b_j$ is contained in every slot and they are identical. As shown in Fig. 4, when a record in a cell is placed, it randomly selects a slot as a destination, then $F_{DAP}$ is performed in the slot. Thus, the $\omega$ value is the number of records in a cell dived by the number of slots. For simplify the introduction, in the previous and rest paper, Probery has only one slot.

Data placement in a block is straightforward. As we have explained, a block may contains record from all cells, and trunks are files to store these records. To reduce the scan scope, trunks in a block should distinguish the records from different cells. However, the strategy is difficult to be implemented. Due to the numerous cells, there are too many trunks and most of them are small files if a trunk is only stores records in one cell. Probery adopts another strategy to place records among trunks. Firstly, trunks of a block are distributed uniformly among nodes to ensure the query parallelism; secondly, records is sequentially written to the trunks line by line while the cell index is treated as the row identifier of each line and stored as its header. Therefore, when trunks are scanned in a query processing, they are scanned in parallel and only the headers of lines are examined to determine whether the lines belong to the queried cells.

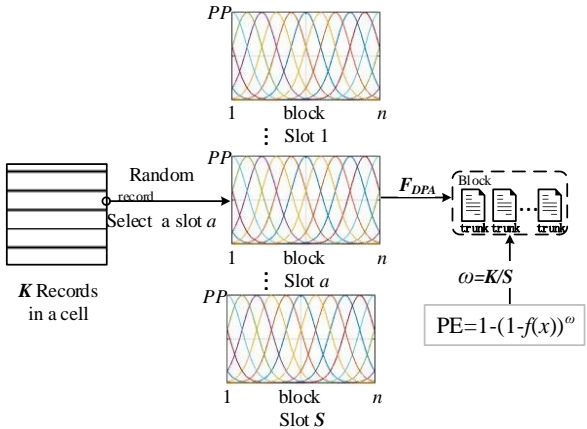

Fig 4 Slots in data placement.

Probery adopts HDFS as the storage. In HDFS, Probery blocks are modeled as HDFS files, and Probery trunks are modeled as HDFS blocks. In HDFS, the data placement is the distribution state of data among nodes. It is for improving both parallelism and I/O efficiency after a dataset is horizontally partitioned into HDFS blocks. Each HDFS block has three replicas, and two replicas are placed on two different nodes in the local rack, the third one is placed on a node in a remote rack. The details of the data placement algorithm for HDFS blocks are explained in our previous work [43].

## 5 QUERY

In this section, we describe the query processing method of Probery. Generally, query is a complex process. It included many operations such as scan, aggregation, join, sort, sub-selection, etc. Probery optimizes the scan operation only, and the other operations rely on the host system. In this section, the host system of Probery is "SQL on Hadoop" because Probery defines a simple SQL-Like clause to query data with a PC confidence. Given a query clause, Probery utilizes a heuristic selection algorithm to scan data and calculates the query error to satisfy the required PC confidence. The other important operations in query processing inherit from host system.

### 5.1 Probability query

Probability query is to query data with a certain confidence of PC. The first step of probability query is to parse the query clause, to determine the queried dimensions and segments, and to retrieve cells in table space where the queried records is contained. The second step is to select blocks using heuristic query algorithm according the given confidence. The third step is to scan trunks in selected blocks, and to retrieve the records that matches the query conditions.

**Definition 9 Query in Probery.** A query in Probery includes three parts: targets, conditions and a confidence. Targets are queried attributes; conditions are query attributes and query ranges; a confidence is the lower bound of PC. A query in Probery is denoted as SQL-Like clauses, it adds a "*with*" clause to specify the confidence. For example:

**select** *targets* **from** *table_name*
**where** *conditions*
**with** *confidence*

Given SQL-Like clauses, Probery parses them to retrieve table names, target attributes, query attributes, query ranges, and the confidence. Each table refers to a table space, in which matched dimensions and segments are located according to the conditions, and then the corresponding cells are selected. For each cell, blocks are selected using Heuristic blocks-selection algorithm (H-selection) whose closure is the probability of query completeness being larger than the specified confidence. Finally, trunks in the selected blocks are traversed to retrieve results.

**Heuristic blocks-selection (H-selection).** H-selection takes matched cells $\{c_i\}$ and confidence $p_0$ as the inputs, and selected blocks $\{b_j\}$ as the output. H-selection ensures the query randomization, namely that the H-selection select blocks randomly even if the query is executed many times with the same confidence, and the query results of them are also different. H-selection ensures that the blocks with tiny PE has change to be queried, but not being skipped every time. Therefore, when blocks are selected, both PE and randomization are considered.

For simplifying the description, we take the first cell as the queried one. Let block $x$, randomly selected from all blocks $B$ for, be the first block for consideration. According to the section 4, PE of block $x$ is $g(x)$, then its PNE (definition 7), which is possibility of not existence, is $1-g(x)$, and denoted as $pne_x$. If $pne_x$ is larger than the confidence $p_0$, it means that the query on block $x$ does not benefit the query completeness because the possibility of "records of the cell being in other blocks", which is also $pne_x$, is already larger than $p_0$. Then $x$ is skipped, meanwhile, the confidence is update to $p_0/pne_x$ for the following deduction:

$$PC[Ept\,|\,(B\text{-}x)] = \frac{PC[Ept\,(B \cup (B\text{-}x))\,]}{PC(B\text{-}x)} = \frac{PC[Ept]}{1\text{-}PE(x)} = p_0/(1\text{-}g(x))$$

where $PC[Ept\,|\,(B\text{-}x)]$ is the expected PC in condition of skipping $x$, the numerator PC[Ept ($B \cup (B-x)$)] is expected PC of querying both with x ($B$) and without $x$ ($B-x$), that is, PC of querying $B$, and denominator is PC of query block $B$ without $x$.

Otherwise, $x$ is not skipped are put back to $B$, Afterward, new block is randomly selected from the rest blocks repeatedly until the updated confidence is equal or very close to 1. The H-selection algorithm is shown as algorithm 1.

ALGORITHM 1 HEURISTIC BLOCKS-SELECTION

| | | |
|---|---|---|
| input: | All blocks $B$ as a set, initial confidence $p_0$ | |
| output: | The select blocks | |
| H-selection | | // the confidence is not close to 1 |
| 1 While !$p_0$≈1 And $B \neq \emptyset$ | | // or some blocks are not checked |
| 2   $x = randomBetween(0, B.size())$ | | //select a block randomly |
| 3   $B.remove(x)$; | | // $x$ will not be selected again |
| 4   If (1- $g(x) > p_0$) | | // PNE, "not exist" probability is // larger than the confidence |
| 5     $p_0 = p_0/(1-g(x))$ | | // skipped, confidence is updated |
| 6   End if | | |
| 7   Else | | |
| 8     $Tmp.add(x)$ | | //block $x$ is NOT skipped |
| 9   End Else | | |
| 10 End while | | //remained blocks in $B$ and $Tmp$ |
| 11 Return $B'=B \cup Tmp$ | | // is selected |

In practice, Probery has a forced closure mechanism to avoid that too many blocks are skipped if confidence is set to be zero or an extreme small value.

### 5.2 Query Error

In the H-selection, we remove the blocks whose PEs are less than the confidence (blocks with small PE), but not select blocks whose PEs are larger than the confidence (blocks with larger PEs) only. The reasons of such strategy are shown as the follows: (1) According to the design of $g(x)$, blocks with larger PEs are less and concentrated, while blocks with small PEs are more and dispersive, therefore, a lot of blocks are skipped in the query processing if the confidence is relatively larger. (2) Because the blocks with small PEs are plentiful, H-selection performed among them has a better randomization than performed among the blocks with larger PEs.

After H-selection, let $B \cup B'$ be the set of select blocks, the joint PE of the selected blocks in the set, is the expected PC, denoted as $p_0'$. There is a query error if $p_0' \neq p_0$. Probery defines the negative error if $p_0' < p_0$ and the positive error if $p_0' > p_0$. Due to the closure conditions shown in line 1 algorithm 1, Probery guarantees negative errors will never happen if we can prove the $p_0 \approx 1$ is satisfied before all blocks is selected.

**Proposition:** In H-selection, $p_0 \approx 1$ is satisfied before all blocks are checked.

**Proof:** Let B is all blocks, $\varepsilon$ is an enough small value.

If $\exists x \in B$ makes $g(x)=(1-p_0)-\varepsilon$, then $1- g(x) = p_0+\varepsilon > p_0$, $x$ is skipped. Afterwards, $p_0$ is updated to $\frac{p_0}{p_0+\varepsilon}$, and it is close to 1.

If !$\exists x \in B$ makes $g(x)=1-p_0-\varepsilon$, it means that $p_0$ is too small that $\forall x \in B$, $1>1-g(x) > p_0 > 0$, then every block is possible to be skipped. Let $p_0=\varepsilon$, and it is updated to $\frac{p_0}{\prod\limits_{x=1}^{n}(1-g(x))}$

$> \frac{\varepsilon}{\varepsilon^n} >1$ if all blocks are skipped. It means $p_0$ increase from $\varepsilon$ to 1 when blocks are skipped.

**Definition 9 Query Error.** Let $B'$ be the selected blocks, $p_0$ be the specified confidence, and $p_0'=1- \prod\limits_{x \in B \cup B'}(1-g(x))$ is the expected PC. The query error $\Delta$ is $p_0'-p_0$. If $\Delta>0$, it is positive error, otherwise it is negative error. Probery guarantees the errors are positive errors only. Larger positive errors reduce the performance promotion while benefit the query completeness.

If we execute the same query with same confidence repeatedly and check the completeness of query result sets one by one, then the true PC, which the proportions of complete results set, is concluded by statistical analysis. Query errors between confidence and true PC are evaluated in the section 7 and proved positive.

## 6 SYSTEM ARCHITECTURE

In this section, the software architecture of Probery implementation is explained. Probery is a key-value oriented query optimization approach, and its implementation required two host systems: a query engine and a distributed file system. Upon them, as shown in Fig. 5, several modules collaborate to implement the table space, probability based data placement and query. The modules include *DrillProxy*, *TableSpace*, *BlockMeta*, *Placement*, *Segmentation* and *ProQuery*. The source code of Probery is available on https://github.com/CloudLab-NEU/Probery.

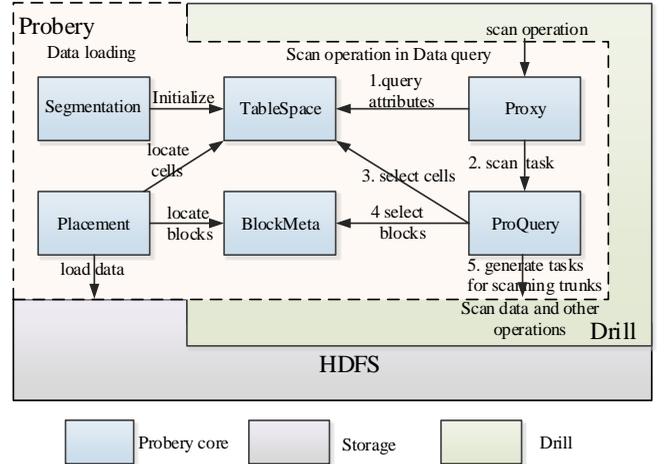

Fig. 5. Software architecture of Probery

- **TableSpace**: The *TableSpace* is a high dimensional data model that is implemented by an in-memory data structure and its access approaches on the master node.
- **Segmentation**: Base on the prior knowledge, *Segmentation* defines query attributes and divides them into near-equal-frequency intervals according to its value range.
- **DrillProxy**: When *DrillProxy* receives query from Drill, it first validates them by referring to *TableSpace*, submits

the query tasks to *ProQuery*.
- **BlockMeta**: *BlockMeta* calculates and store the PPs and PEs of all blocks. It is a metadata of files in the *Storage*.
- **ProQuery**: *ProQuery* is the core of Probery that implements strategies and algorithms in section 5.
- **Placement**: *Placement* is the core of Probery. Referring to the *BlockMeta*, it places the data into chunks of blocks, as described in section 4.
- **Query Engine** (**Drill**). Probery technically supports other "SQL on Hadoop" query engine but currently we only implement the Drill-based one.
- **Storage**. Key-values data is stored in a distributed file system, currently only HDFS is supported.

## 7 EXPERIMENTS

In this section, the three critical features of Probery, namely data placement, query completeness and query speedup, are validated, and query performance of Probery is compared with other "SQL on Hadoop" systems in a cluster.

### 7.1 Query Validation
**(1) Setup**

**Scope**. Validate the effectiveness and efficiency of DPA and probability query in a standalone environment.

**Experiment environment.** We adopt a standalone server, because only the data placement and distribution, confidence, and query efficiency are validated, but not the performance

**Selection of competitors.** We compare the DPA time, balance of data distribution, observer values and the expected values of PC, and query efficiency in various conditions.

**Experimental data.** We adopt the generated dataset. There are maximum 1000 records in a trunk. We set the number of blocks be 2000 ($B_1$), 4000 ($B_2$), 6000 ($B_3$), 8000 ($B_4$) and 10000 ($B_5$), and the scale of the datasets be 20 ($D_1$), 40 ($D_2$), 60 ($D_3$), 80 ($D_4$) and 100 ($D_5$) million records. So, blocks averagely contain at least 10,000 records (10 trunks) and at most 50,000 records (50 trunks) if the data distribution among blocks is uniformly. Each record contains three attributes, denoted as *KeyA*, *KeyB*, and *KeyC*. Attributes are all dimensions, and the number of segments on each dimension is as same as the number of blocks. The values of keys are random integers from 0 to $10^8$.

**Experimental cases.** All attributes in the dataset are selected, and the query conditions are ranges defined by a random segment on each dimension in a query. The confidences are from 0.1 to 0.9, and denoted as $C_1$ to $C_9$, while observed PC of a specific confidence is summarized using 1000 queries.

**Symbols.** There are a few symbols involved in the experiment analysis, which are introduced in above illustrations. We list the symbols in **Error! Reference source not found.**2.

TABLE 2. SYMBOLS USED IN EXPERIMENT ILLUSTRATION

| Symbol | Description |
|---|---|
| $B_iD_jC_k$ | A test case that refers to the query on $B_j$ number of blocks, $D_j$ dataset and specified confidence $C_k$. $i$ and $j$ are in {1,2,…,5} and $k$ is in {1,2,…,9}. |
| $opc(B_iD_jC_k)$ | The observed query completeness of the case $B_iD_jC_k$. Probery expects that it is larger but close to $k/10$. |
| $qe(B_iD_jC_k)$ | The query efficiency (QE) of $B_iD_jC_k$, which means the ratio between the number of matched blocks and searched blocks. QE represents the speedup rate. |

**(2) DPA in loading**

The loading performance of Probery includes DPA, namely that choose the target block for a record, and the stage of writing data to the disk. We focus on DPA performance since the latter is the same for any loading strategy. We expect that Probery consumes less time in selecting the target block when data is loaded.

Fig. 6 shows the average PDA time of various block numbers and data volumes. Overall, the time consumption of DPA is no more than 5 seconds for 100 million records. On one hand, the DPA does not bring much cost to data loading comparing with the cost of I/O operations. For example, if the data loading is performed by two threads, one of which takes the responsibility of placing data and the other takes the responsibility of loading data, then the former thread never blocks the latter one. Therefore, the data loading performance is dominated by I/O performance. The cost of DPA is ignorable. On the other hand, data loading is executed only once, and all queries benefit from the new functionality, therefore the proportional cost is very low. In conclusion, the extra cost of DPA in data loading is negligible.

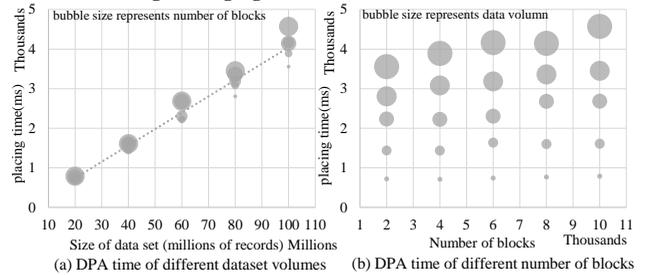

Fig. 6. Comparison of DPA time of different cases

Fig. 6 shows that there was a significant linear and positive correlation between time consumption linearly and data volume. For same data volume, the time consumption of DPA slightly increases with the number of blocks because Probery takes more efforts to maps probabilities to block indexes. In the Java implementation, we do not adopts *org.apache.commons.math3.distribution*. *NormalDistribution* to calculate probability or cumulative probability due to the performance reasons. Instead of that, a lookup table is initialized and $f^1(x)$ is implemented as binary search on the table. When there are more blocks, the table is larger, and the binary search is a little bit costly.

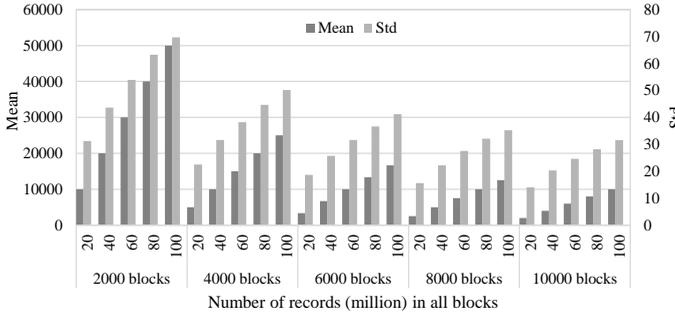

To validate that the data distributed in each block is balance, we calculate the mean and standard deviation of data volumes in blocks under the 25 combinations of $B_iD_j$ and show the results in Fig. 7. The mean values, denoted as dark bars and primary axis, equal to the data volume divided by block number. The standard deviations, denoted as orange bars and secondary axis, are positively correlated to the mean values but no larger than 80. We believe that the data volume distributed in blocks are balance.

Fig. 7. The data distribution among blocks (mean and standard deviation)

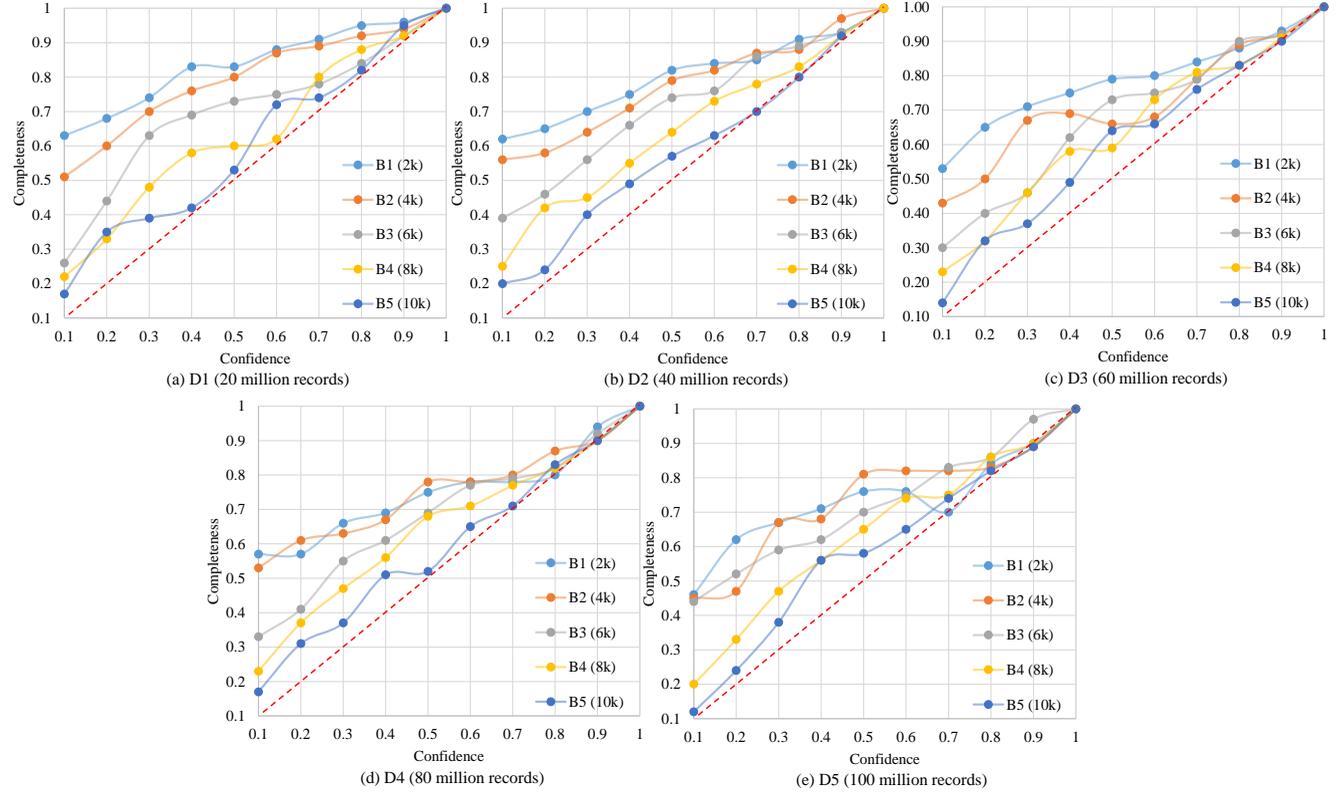

Fig. 8. Relationship between observed PCs and confidences in various cases

**(3) Query completeness**

Being the critical functions, the probability of query completeness is validated in the experiments. We expect the observed PC, $opc(B_iD_jC_k)$ of Probery is larger but close to $k/10$ ($k$ is subscript of $C$). Fig. 8 shows the 50*5 observed PCs that concluded from 1000 random queries and grouped by $D_j$ first and $B_i$ later. From Fig. 8 the following conclusions are drawn:

(1) For all cases, the observed PCs are no smaller than the specified confidences, because each curve, which represent the relationship between observed PC ($y$) and confidences ($x$), is not below the reference line ($y=x$).

(2) For the most cases, the observed PC increases with the confidence, because the curves are monotonically increasing. However, there is some exceptions, such as $B_2D_3$ in Fig.8-(c), and $B_5D_5$ in Fig. 8-(e). Because the H-selection algorithm selects blocks randomly, sometime the floating of PE causes the situation that the observed PC for confidence $p_a$ is even larger than that for the confidence $p_b$ despite $p_a$ is slightly smaller than $p_b$.

(3) Obviously, the observed PCs are not exactly obeying the confidences. For all cases, the curves end at the range (1, 1), but start from the range (0.1, 0.6), that is to say, when confidence increases from 0.1 to 1, the same as the observed PC in some cases, while the observed PC increases slowly from 0.5 to 1 in the other cases. Where the curves start relates to the number of blocks, but not the data volume. If there are more blocks, the PEs of them are closer, namely that the granularity of PE distributed on these blocks are finer, hence H-selection has more change to select the blocks with different PEs. On the contrary, PEs would concentrate on few distinct values if there are few blocks, then the observed PCs are close no matter what the specified confidences are.

(4) The best cases that satisfies the design purposes

of Probery are $B_5D_j$, because their curves close to the reference line ($y=x$) in each sub-figure mostly. We believe the $f(x)$, where $\sigma=0.3989$ and $\mu=2$, adapts $B_5$ well. Technically, we could also define $f(x)$ with different $\mu$ to adapt $B_1$ to $B_4$ better. However, $f(x)$ with larger $\mu$ rises a problem of "too many 0s in $g(x)$", meanwhile $f(x)$ with smaller $\mu$ rises a problem of "values of $g(x)$ are similar", as explained in Fig 9. In practice, the number of blocks is fixed so that a carefully designed $f(x)$ is beneficial.

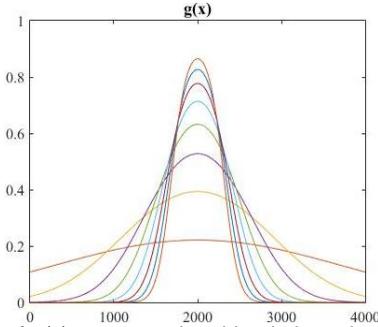

Fig 9 Image of $g(x)$, curves ordered by their peaks are $\mu=0.5$, 1, 1.5, 2, 2.5, 3, 3.5, 4, where $\omega=1000$, $n=4000$, $\sigma=0.3989$,

**(4) Query efficiency (QE)**

Probery trade off the possibility of query completeness for query performance. In this experiment, we define the query efficiency (QE), the ratio between matched blocks and searched blocks, to indicate the performance speedup when confidence decrease. The theoretical maximum QE is 1 but it can never be reached. We expect that $qe(B_iD_jC_x) > qe(B_iD_jC_y)$ if $x < y$. There are 250 combinations of test cases, and we summarize the five-statistical numbers (min, Q1, median, Q2, max) for 1000 queries of each case, and draw the boxplot as shown in Fig 10. From Fig 10 the following conclusions are drawn.

(1) For all cases, QEs of probability query are definitely larger than non-probability query (confidence=1), comparing Fig. 10-(a), (b), (c), (d), (e) to Fig. 10-(f). While for each case with same $B_iD_j$, QE decreases with in the creasing confidence because less blocks are skipped when confidence is larger. However, QEs does not proportional decrease as confidences increase because almost half blocks whose PEs are neither 0 nor 1 (see Fig.9 $u=4$), thus the H-selection is performed on rest blocks.

(2) For the same number of block, QE increase with data volume, namely, $qe(B_iD_x) > qe(B_iD_y)$ if $x>y$. For the same data volume, QE increase with the decreasing number of block, namely, $qe(B_xD_j) > qe(B_yD_j)$ if $x<y$. These regularities are also satisfied when confidence=1, as shown in Fig. 10-(f). In this experiment, the QE fluctuation is dominated by the queries themselves but not any strategies of Probery. As explained in section 7.1, the query condition is a random segment on each dimension, and the number of segments is equal to the number of blocks. When the latter is fixed, the more data volume is, the more records a block contains, and the more blocks records in a segment spreading to, thus the matched blocks increases, so the same with QE. When the data volume is fixed, less blocks also cases records in a segment spreading more blocks, thus QE also increases.

(3) In fact, the number of block or data volume do has a tiny effects on QE through "placement count", which is $\omega$ in equation $g(x)=1-(1-f(x))^\omega$. When more records are placed on a block, $\omega$ is getting larger, then the curve of $g(x)$ climes from 0 and converges to 1 quickly. In such situation, QE increases, and QEs of different confidences are similar due to there are too many 1s and 0s of PEs. The "placement count" is positively related to the block size, which increases with the data volume, and the decreasing number of blocks. However, such effects are insignificant.

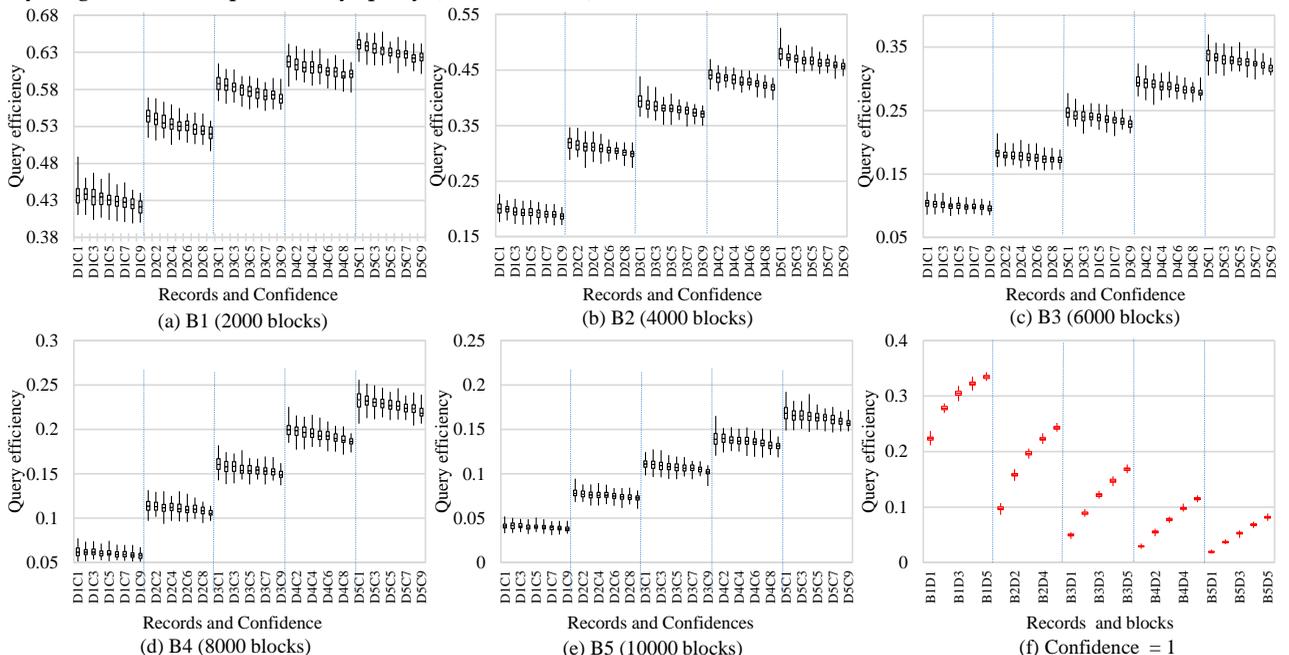

Fig. 10. QE of possible incomplete query cases (a)-(e), comparing with QE of complete query cases (f), namely confidence=1

## 7.3 Performance Comparison

Before comparing Prober with other "SQL on Hadoop" systems, there are two concepts to be clarified: (1) Probery is an optimization approach but not a "SQL on Hadoop" system. Its implementation relies on a host system, such as Drill in this paper. (2) Probery updates the scan operation of the host system. It reduces the scan scope. The other important operations, such as aggregation, join, sort, sub-selection, of the host system remain unchanged. It also inherits the performance features of the host system;

**(1) Setup**

**Scope**. Compare query performance of Probery with other "SQL on Hadoop" systems in a cluster. The scalable parameter is result-set size and dataset size. The scalability of nodes are not included because Probery does not bring changes to the host platform.

**Experiment environment.** We run experiments on a cluster with 12 physical machines. Each node has the same specification, i.e., Intel Core i7, 1TB hard disk, 8 GB memory, CentOS 7, and moderate I/O performance. The gigabit Ethernet was connected using a Dell PowerConnect 5548.

**Selection of competitors.** We select Apache Drill as the query engine of Probery, because MapReduce is a little bit old-fashioned. We set Probery with different confidences, namely 1, 0.8, 0.5 and 0.2, and denoted as Probery1, Probery0.8, Probery0.5, Probery0.2, respectively. We compare the query performance of them with Drill, Impala and Hive on Tez. They are all "SQL on Hadoop" system [44], but the former two base on MPP and Apache Parquet [45], and the latter one bases on DAG framework [46] and ORC [47]. We treat Impala and Hive as two baselines of performance. Generally, Impala is faster than Drill due to its high I/O performance [48], and Hive-Tez is faster than Hive-MR, and what more, inline indexes in ORC files make Hive even more effective. We plan to see whether Probery is comparable to Impala as the efficient I/O and memory solution, and to Hive as a sophisticated indexes solution.

**Experimental data.** We adopt table *UserVisits* in Web Data Analytics (WDA) micro benchmark [48]. The schema of *UserVisits* are:

| | |
|---|---|
| *sourceIP* VARCHAR(116), | *destURL* VARCHAR(100) |
| *visitDate* DATE, | *adRevenue* FLOAT, |
| *userAgent* VARCHAR(256) | *countryCode* CHAR(3) |
| *languageCode* CHAR(6) | *searchWord* VARCHAR(32) |
| *duration* INT | |

We design five scales of dataset, $S_1$ to $S_5$, the data volume are 5GB, 10GB, 20GB, 40GB and 80GB, respectively. The *visitDate*, *adRevenue*, and *sourceIP* fields are picked uniformly at random from specific ranges, modeled as query attribute, and divided into 10 segments. In the table space, there are 1000 cells. All other fields are picked uniformly from sampling real-world data sets.

**Query cases.** We only design simple query cases without aggregation, join, sub-selection and sort, because these are not our optimization targets, also we plan to highlight the scan performance, otherwise the performance different is hidden by these mechanism. We randomly set the query condition on each query attribute and make sure that the expected number of segments, from 1 to 7, on each dimension of the table space are matched, and there is no more queries in a cell. Due to the uniform distribution of query attributes, the query scale, as the ratio between size result-set size and dataset, is $(w/10)^3$ if $w$ segments are selected on each query attributes. In the experiment, query scales are 0.001 ($w$=1), 0.008 ($w$=2), 0.027 ($w$=3), 0.064 ($w$=4), 0.125 ($w$=5), 0.216 ($w$=6), 0.343 ($w$=7). The confidences are 1, 0.8, 0.5 and 0.2, as competitors called Probery1, Probery0.8, Probery0.5, Probery0.2, respectively. Beside, Hive sort one of the queried attribute, *adRevenue*, for fully utilizing the inline index of OCR files.

**(2) Query Scales**

We set Probery for different confidence, denoted as Probery1, Probery0.8, Probery0.5 and Probery0.2. We compare the query performance of them with Drill, Impala and Hive under the different query scales, to evaluate the parallelism of scan operation.

The simple query case contains three operations only: scan, filtering and output. The query time is dominated by the scan operation. Impala, Hive and Drill perform a full scan of the *UserVisits* and apply the filter to generate the result set, and Probery does the same because Probery bases on Drill, and records in a cell is possible to place to any block. Due to the systems adopt full-scan, the scan time is only affected by parallelism, and the parallelism is affected by the data balance. During the full-scan of Hive, ORC's Predicate Pushdown consults the inline indexes to try to identify when entire blocks can be skipped all at once, however, it means Hive will scan files faster, but not scan less files. Therefore, we expect that query time is stable for different query scale in Drill, Impala, Hive and Probery1. For Probery0.x, there are some blocks being skipped. The query time should be stable because the skipped blocks will not affects the parallelism if each blocks is distributed to nodes uniformly.

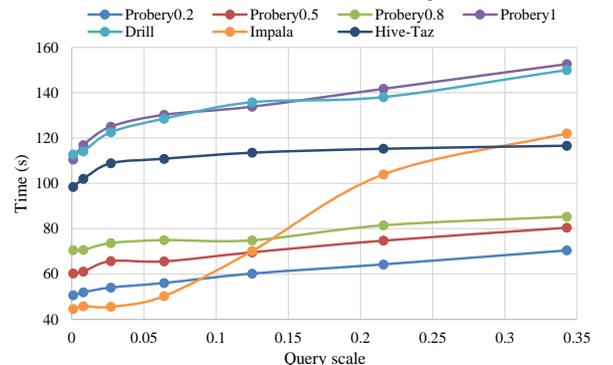

Fig. 11. Query time on different query scales

Fig. 11 shows the query time on different query scales for the competitors. The curves are all stable except the Impala one. Impala becomes bottlenecked on the ability to persist the results back to disk when they are huge. The experimental results prove our ex-pectations. Probery distributed chunks in a cell uniformly among nodes, otherwise, the parallelism is changed if blocks are skipped, and the query time is affected.

The scan time occupy about 90% the total query time in Probery, Drill and Impala, and about 85% in Hive. When more records in the query results, the output operation takes more time, it can explain why curves are slightly upward. The experiment is under the 80 GB dataset. The overall performances show that Probery1 is same with Drill, and close to Hive, also the Prober0.x and Impala are comparable. We will explain them next.

**(3) Dataset Scale**

In this experiment, we compare the performance and scalability of Probery to Drill, Impala and Hive. We expect two results: first, query performance and scalability of Probery1 and Drill are the same because Probery bases on Apache Drill, and it scan all blocks when confidence=1. And Probery0.x are expected to be faster than Impala, a well-known high-performance query engine, and Hive, a system with sophisticated inline index and the sorted attributes. Fig. 12 shows the experimental results and conclusions are drawn as the follows:

(1) As the expected, performance of Probery0.8, Probery0.5 and Probery0.2 are gradually improved, while performance of Probery1 and Drill are almost the same. However, for all cases, Probery0.x is averagely 1.8x faster than Drill, 1.3x faster than Impala and 1.6x faster than Hive. The optimization effects of Probery is obvious.

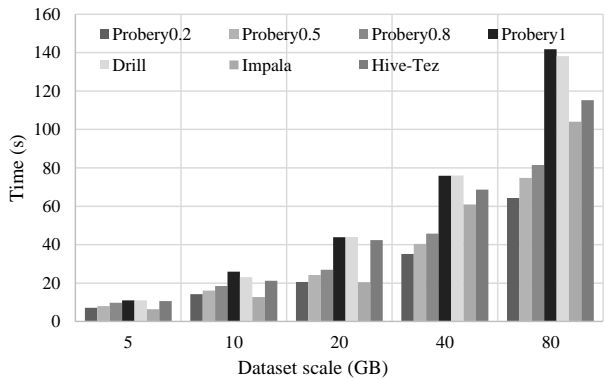

Fig. 12. Query time on five datasets, the bars represent Probery0.2, Probery0.5, Probery08, Probery1, Drill, Impala and Hive-Tez from the left to the right Impala

(2) Probery0.x cannot be more efficient than Impala for the former three datasets, but still competitive. Moreover, Probery0.x wins in the 40GB and 80GB dataset. Comparing with the optimization approaches of Impala, such as distributed plan tree, I/O scheduling and memory usage, Probery trade off the possibility of query completeness for performance. Given the same query, the query results become huge when data set is huge. Impala becomes bottlenecked on persisting the results back to disk.

(3) Probery has sub-linear scalability on dataset volume, and such feature is inherited from Drill. Impala shows very stable performance on 5GB, 10GB and 20GB datasets, but sharply turn worse on 40 GB and 80 GB datasets because of memory limitation. Probery does not rely on the pre-computation or materialized techniques. Its in-memory data structure, table spaces, contians the metadata of dimension and segment only, and the mapping relationship between cells and blocks is calculated but not lookupped, for reasons above, the memory is saved. So Probery0.2 is slight faster than Impala on 20GB dataset, and Probery0.x catch up Impala for the 40GB and 80GB datasets. Hive, with its inline index, also has sub-linear scalability on dataset volume.

(4) Comparing with the Impala's memory solution and Hive's inline index solution, Probery meet the design goals. It successfully trade off PC for query performance. Beside, on one hand, Probery is an approach working along with Drill and optimizations of Drill also improve Probery; on the other hand, Probery also can implementation over other SQL-on-Hadoop systems, to overlap its optimization to the systems.

# 8 CONCLUSION

This paper presents the design, implementation, and evaluation of Probery, a key-value store supporting probability-based query approach, in which the probability means the confidence of query completeness. Probery is built on a Hadoop HDFS, and Drill as query engine. It theoretically support other query engines such as SparkSQL. We proposed the following models and algorithms for Probery:

(1) Table spaces, dimensions, segments, cells, blocks and trunks, which are the elements of multi-dimensional data model as the logical schema, and data containers as the physical schema of the key-value pairs, are the proper abstraction for a key-value store.

(2) The probability of query completeness (PC), probability of existence (PE) and probability of placement (PP) address the probability-based query of Probery well, also show the differences with other data oriented probability query solutions.

(3) The data placement algorithm ensures both the uneven probability distribution and balance data distribution, and query parallelism. PDA has better extensibility; no special efforts are required if new data or new nodes are introduced.

(4) SQL-Like statements adding "*with*" clause to specify the confidence simplifying the front-end queries. The H-selection algorithm in the probability query skips the blocks with lower PE.

(5) The system architecture and components of Probery implementation.

The experimental results demonstrated that DPA in loading process is efficient, the query completeness and query efficiency meet the design purposes. For the Drill based Probery implementation, its query performance is the most stable and efficient comparing with that of Drill and Hive, and it is competitive or even better when data volume is huge, comparing with that of Impala.

Notice that the definition of *f(x)* affects the efficiency of Probery, and there may exits a more sophisticated one beyond our knowledge. Nevertheless, *f(x)* in Probery is replaceable, and changing a new function will not affects the other modules, in addition, we think about add an *f(x)*-evaluation module to Probery. In future studies, Probery could be further optimized as follows: more investigation on *f*(x) in DPA because the optimization effect is dominated by a well-designed *g*(x); more sophisticated data structures for the table space and storage format for the chunk; also adapting Probery to other fashionable query engines.

## ACKNOWLEDGMENT

This paper is supported by the National Natural Science Foundation of China under Grant No. 61672143, 61433008, U1435216, 61662057, 61502090, and 61402090, also the Fundamental Research Funds for the Central Universities under Grant No. N161602003. The authors thanks Wu Jingbo for initial works on Probery.

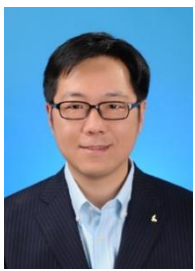

**Jie Song** received the Ph.D. degree from Northeastern University in 2010. He is Associate Professor of Software College, Northeastern University. His research interest includes big data management and machine learning.

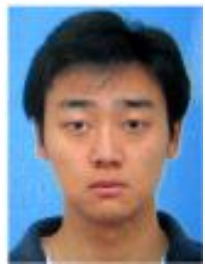

**Yichuan Zhang** received the Ph.D. degree from Northeastern University in 2010. He is a Lecture of Software College, Northeastern University. His research interest includes big data management and iterative computing.

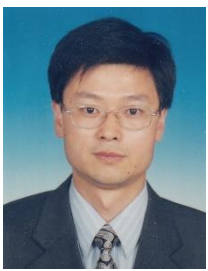

**Yubing Bao** received the Ph.D. degree from Northeastern University in 2003. He is a Professor of Computer Science and Engineering of Northeastern University. His research interest includes big data management and iterative computing.

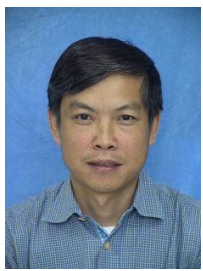

**Ge Yu** received his Ph.D. degree in computer science from Kyushu University of Japan in 1996. He is a Professor of Computer Science and Engineering, Northeastern University. His research interest includes database theory and technology, distributed and parallel systems.